\documentclass[aps,twocolumn,groupedaddress,showpacs]{revtex4}
\usepackage{graphicx}
\newcommand\beq{\begin{eqnarray}}
\newcommand\eeq{\end{eqnarray}}
\newcommand\qom{\frac{q}{M}}

\newcommand\qovm{\frac{q}{2M}}
\newcommand\fip{\frac{\mid \phi_{\mu}(0) \mid^{2}}{4\pi}}

\newcommand\vlsr{C_{V}^{L}C_{S}^{R*}}

\newcommand\pmo{{\bf |P_{\mu}|}}
\begin{document}
\title{Right-handed Neutrinos in \mbox{\boldmath$\nu_{\mu} + e^{-}\rightarrow
\nu_{\mu} + e^{-}$} Elastic Scattering}
\author{W. Sobk\'ow}
\email{sobkow@rose.ift.uni.wroc.pl}
\affiliation{Institute of Theoretical Physics, University of Wroc\l{}aw,
Pl. M. Born 9, PL-50-204~Wroc\l{}aw, Poland}
\date{\today}
\begin{abstract}
In this paper a scenario admitting the participation  of the
exotic right-handed  scalar $g_{S}^{R}$  coupling in addition to
the standard left-handed  $g_{V}^{L}, g_{A}^{L}$ couplings in the
low-energy $\nu_{\mu} + e^{-}\rightarrow \nu_{\mu} + e^{-} $
scattering is considered. The main  goal is to show how the
presence of the right-handed neutrinos in the $(\nu_{\mu}e^{-})$
scattering changes the laboratory differential cross section in
relation to the Standard Model prediction. The $ (\nu_{\mu}
e^{-})$ scattering is studied at the level of the four-fermion
point interaction. Muon-neutrinos are assumed to be polarized
Dirac fermions and to be massive. In the laboratory differential
cross section the new interference term between the standard
$g_{V}^{L}$ coupling of the left-handed neutrino and exotic
$g_{S}^{R}$ coupling of the right-handed neutrino  appears which
does not vanish in the limit of massless neutrino. This additional
contribution, including information about the transverse
components of neutrino polarization, generates the azimuthal
asymmetry in the angular distribution of the recoil electrons.
This regularity would be the proof of the participation of the
right-handed neutrinos in the $(\nu_{\mu} e^{-})$ scattering.
\end{abstract}
\maketitle
The $(V-A)$ structure  of weak interactions at low energies describes
only what  has been measured so
far. We mean here  the measurement of the electron helicity
\cite{Bob}, the indirect measurement of the neutrino helicity
\cite{Gold}, the asymmetry in the distribution of the electrons
from $\beta$-decay \cite{CWu} and
the experiment with muon decay \cite{Gar} which
confirmed parity violation \cite{Lee}.
Feynman, Gell-Mann and
independently Sudarshan, Marshak \cite{Gell} established that
only left-handed vector $ V$, axial $ A$ couplings can take part
in weak interactions because this yields the maximum symmetry
breaking under space inversion, under charge conjugation; the
two-component neutrino theory of negative helicity;  the
conservation of the combined symmetry $ CP$ and of the lepton
number.
In consequence it led to the conclusion that produced neutrinos
in $V-A$ interaction can only be left-handed. However Wu
\cite{SWu} indicated  that  both  standard left-handed $(V,
A)_{L}$ couplings and exotic right-handed $(S, T, P)_{R}$
couplings may be responsible for the negative electron helicity
observed in $\beta$-decay. It would mean that generated neutrinos
in $(S, T, P)$ interactions  may also be {\it right-handed}.
Recent tests do not provide a unique
 answer as to the presence of  the exotic weak interactions.
So Shimizu {\it et al.}  \cite{Shimizu} determined the ratio of the
strengths of scalar and tensor couplings to the standard vector coupling
in $K^{+}\rightarrow \pi^{0}+e^{+}+\nu_{e}$ decay at rest
assuming the only left-handed neutrinos for all interactions.
Their results indicated the
compatibility with the Standard Model (SM) \cite{Glashow,Wein,Salam} prediction.
Bodek {\it et al.}  at the PSI \cite{Bodek}  looked for the
evidence of the violation of time reversal invariance measuring
$T$-odd transverse components of the positron polarization in
$\mu^{+}$-decay. They also admitted the presence of the only
left-handed neutrinos produced by the scalar interaction. The
recent results presented by the DELPHI Collaboration \cite{Delphi}
concerning the measurement of the Michel parameters and the
neutrino helicity in $\tau$ lepton decays indicated the
consistency with the standard $V-A$ structure of the charged
current weak interaction. However on the other hand, the achieved
precision of measurements still admits the deviation from the
pure $V-A$ interaction, i.e. the possible participation of the
exotic interactions with the right-handed neutrinos beyond the
SM. It is necessary to carry out the new  high-precision tests of
 the  Lorentz structure and of the handedness structure of the
weak interactions at low energies in which the {\it components of
the neutrino polarization} would be measured, because in the
conventional observables  the interference contributions coming
from  the right-handed neutrinos vanish in the  limit of massless
neutrino \cite{Wolf,Miran,Sobkow}.
Frauenfelder et al. \cite{spin} pointed out that one has to measure either the neutrino
polarization (spin) or the neutrino-electron correlations to determine
the full Lorentz structure of the weak interactions.
\par The main our goal is to show how the presence of
the right-handed neutrinos in the $(\nu_{\mu}e^{-})$ scattering
changes laboratory differential cross section in relation to the
SM prediction with the left-handed neutrinos. We will consider
the process of the $(\nu_{\mu}e^{-})$   scattering at the level
of the four-fermion point (contact) interaction. Muon-neutrinos
are assumed  to be  massive Dirac fermions and to be polarized.
In these considerations, the incoming neutrinos come
from the muon-capture, where the production plane is spanned by
the initial muon polarization and the outgoing neutrino momentum.
However in practice, the suitable low-energy polarized neutrino
source could be the strong chromium source $(^{51}Cr)$.
Because there exist the models of the spontaneous symmetry
breaking under time reversal \cite{TDLee} in which the scalar weak
interaction can appear, the possible participation of the exotic
right-handed scalar $S$ coupling in addition to the standard
left-handed vector $V$ and axial $A$ couplings is admitted. It
means that in the $V-A$ interaction, the incoming muon-neutrinos
are always left-handed, but in the scalar $S$ interaction the
initial neutrinos are right-handed (the outgoing neutrinos are
left-handed). The couplings constants are denoted as $g_{V}^{L},
g_{A}^{L}$ and $g_{S}^{R}$ respectively to the incoming neutrino
handedness: \beq { \cal M} &=&
\frac{G_{F}}{\sqrt{2}}\{(\overline{u}_{e'}\gamma^{\alpha}(g_{V}^{L}
- g_{A}^{L}\gamma_{5})u_{e}) (\overline{u}_{\nu_{\mu'}}
\gamma_{\alpha}(1 - \gamma_{5})u_{\nu_{\mu}})\nonumber \\
&  & \mbox{} +
\frac{1}{2}g_{S}^{R}(\overline{u}_{e'}u_{e})(\overline{u}_{\nu_{\mu'}}
(1 + \gamma_{5})u_{\nu_{\mu}})\},
 \eeq
 where $ u_{e}$ and  $u_{e'}$
$(u_{\nu_{\mu}}\;$ and $\; u_{\nu_{\mu'}})$ are the Dirac
bispinors of the initial and final electron (neutrino)
respectively, $G_{F}= 1.16639(1)\times 10^{-5}GeV^{-2}$  is the
Fermi coupling constant \cite{Data}. In our considerations the
system of natural units with $\hbar=c=1$, Dirac-Pauli
representation of the $\gamma$-matrices and the $(+, -, -, -)$
metric are used \cite{Mandl}.
 \par Because the carried out investigations for the
$\mu^{-}$-capture \cite{Sobkow} led to the conclusion that the
production of the right-handed neutrinos by the exotic scalar
interaction manifests the non-vanishing  transverse components
of the neutrino polarization in the limit of massless neutrino,
one expects the similar regularity in the $(\nu_{\mu}e^{-})$
scattering. The matrix element for the $\mu^{-}$-capture and
formula on the magnitude of the transverse neutrino  polarization
vector $|\mbox{\boldmath $\eta_{\nu }^{T}$}|$, in the limit
of vanishing neutrino mass, were as follows:
\begin{widetext}
\beq {\cal H_{\mu^-}} & = & C_{V}^{L}({\overline\Psi}_{\nu}
    \gamma_{\lambda}(1 - \gamma_{5})\Psi_{\mu})
    ({\overline\Psi}_{n}\gamma^{\lambda}\Psi_{p}) \\
&  &  \mbox{} + C_{A}^{L}({\overline\Psi}_{\nu}
    i\gamma_{5}\gamma_{\lambda}(1 - \gamma_{5})\Psi_{\mu})
    ({\overline\Psi}_{n}i\gamma^{5}\gamma^{\lambda}\Psi_{p})
 +  C_{S}^{R}({\overline\Psi}_{\nu}(1 - \gamma_{5})\Psi_{\mu})
  ({\overline\Psi}_{n}\Psi_{p}), \nonumber\\
|\mbox{\boldmath $\eta_{\nu }^{T}$}| & = &
\frac{\sqrt{{\bf < S_{\nu}\cdot({\hat
P}_{\mu}\times {\hat q}) >}_{f}^{2} + {\bf < S_{\nu}\cdot {\hat
P}_{\mu}>}_{f}^{2}}}{<\bf 1>_{f}}
 =  \pmo
|\frac{C_{S}^{R}}{C_{V}^{L}}|(1+\qovm) \\
&  & \times \{(3+\qom)|\frac{C_{A}^{L}}{C_{V}^{L}}|^{2} + (1+\qom)
+ |\frac{C_{S}^{R}}{C_{V}^{L}}|^{2}
- \qom|\frac{C_{A}^{L}}{C_{V}^{L}}| cos(\alpha_{AV}^{L})\}^{-1}, \nonumber\\
{\bf <S_{\nu}\cdot({\hat P}_{\mu}\times\hat{q})>}_{f}
  & = & - \fip\pmo (1 + \qovm)Im(\vlsr), \\
{\bf <S_{\nu}\cdot{\hat P}_{\mu}>}_{f}
  & = &  \fip\pmo (1 + \qovm)Re(\vlsr),
\eeq
\end{widetext}
 where $C_{V}^{L}, C_{A}^{L}, C_{S}^{R}$ - the complex coupling constants for the
standard  vector, axial and exotic scalar weak interactions
denoted respectively to the outgoing neutrino handedness;
 ${\bf <S_{\nu}\cdot({\hat P}_{\mu}\times\hat{q})>}_{f},
{\bf <S_{\nu}\cdot{\hat P}_{\mu}>}_{f}, {\bf <1>_{f}}$ - the
$T-$odd and $T-$even transverse components of neutrino
polarization and the probability of muon capture, respectively;
$q, M$ - the value of the  neutrino momentum and the nucleon mass;
$\pmo$ - the value of the muon polarization in $1s$ state;
$\phi_{\mu}(0)$ - the value of the large radial component of the
muon Dirac bispinor for $r=0$; ${\bf \hat{q}, {\hat P}_{\mu}}$ -
the direction of the neutrino momentum and of the muon
polarization, respectively; $\alpha_{AV}^{L} \equiv
\alpha_{A}^{L} - \alpha_{V}^{L} $ - the relative phase between
the standard $C_{A}^{L}$ and  $C_{V}^{L}$ couplings.
\par It can be seen that
the neutrino  observables consist exclusively of the interference
term  between the standard left-handed vector $V_{L}$ coupling
 and exotic right-handed scalar $S_{R}$ coupling.
It can be understood as the interference between the neutrino
waves of negative and positive handedness. If we admit also the
presence of the left-handed scalar $S_{L}$ coupling in addition to
the right-handed exotic scalar $S_{R}$ coupling, we get the
interferences between the $(V, A)_{L}$ and $S_{L}$ couplings
proportional to the neutrino mass, and they vanish in the limit
of massless neutrino.
 Our coupling constants $C_{V,A}^{L},
C_{S}^{R}$ can be expressed by Fetscher's couplings
$g^{\gamma}_{\epsilon \mu}$ for the normal and inverse muon decay
\cite{Data}, assuming the universality of weak interactions. The
induced couplings generated by the dressing of hadrons are
neglected as their presence does not change qualitatively the
conclusions about transverse neutrino polarization. Here,
$\gamma= S, V, T$ indicates a scalar, vector, tensor interaction;
$\epsilon, \mu=L, R$ indicate the chirality of the electron or
muon and the neutrino chiralities are uniquely determined for
given $\gamma, \epsilon, \mu$. We get the following relations:
\beq
C_{V}^{L} &=& A(g_{LL}^{V} + g_{RL}^{V}), \\
-C_{A}^{L} &=& A(g_{LL}^{V} - g_{RL}^{V}), \nonumber\\
C_{S}^{R} &=& A(g_{LR}^{S} + g_{RL}^{S}), \nonumber
\eeq
where $A\equiv(4G_{F}/\sqrt{2})cos\theta_{c}$,
$\theta_{c}$ is the Cabbibo angle.
In this way,  the lower  limits on the $C^{L}_{V,A}$ and upper
limit on the $C_{S}^{R}$  can be calculated, using the current
data \cite{Data}; $|C_{V}^{L}|>0.850 A, \;|C_{A}^{L}|>1.070 A,
\;|C_{S}^{R}|<0.974 A$. In consequence, one gives the upper bound
on the
 magnitude  of the transverse neutrino polarization vector proportional
 to the value of the muon polarization;
 $|\mbox{\boldmath $\eta_{\nu }^{T}$}|\leq 0.334 \pmo $ (for
$\alpha_{AV}^{L}=0$).
 The obtained limit has to be divided by the $\pmo$
 to have the upper bound
on the physical  value of the transverse  neutrino polarization
vector generated by the exotic scalar interaction;
$|\mbox{\boldmath $\eta_{\nu}^{' T}$}| =
|\mbox{\boldmath $\eta_{\nu }^{T}$}|/ \pmo \leq 0.334$.
\par To describe $(\nu_{\mu}e^{-})$ scattering the following
observables are used: \mbox{\boldmath $\eta_{\nu}$} - the full
vector of the initial neutrino polarization in the rest frame,
${\bf q}$ - the incoming neutrino momentum, ${\bf p_{e'}}$ - the
outgoing electron momentum. The laboratory differential cross
section for the $\nu_{\mu}e^{-}$ scattering, in the limit of
vanishing neutrino mass, is of the form:
\begin{widetext}
\beq
\label{przekr}
\lefteqn{\frac{d^{2} \sigma}{d y d \phi_{e'}}
= (\frac{d^{2} \sigma}{d y d \phi_{e'}})_{(V, A) } +
(\frac{d^{2} \sigma}{d y d \phi_{e'}})_{(S)}
+ (\frac{d^{2} \sigma}{d y d \phi_{e'}})_{(V S)},}\\
(\frac{d^{2} \sigma}{d y d \phi_{e'}})_{(V, A)}
&=& B \{ (1-\mbox{\boldmath $\eta_{\nu}$}\cdot\hat{\bf q}
)[(g_{V}^{L} + g_{A}^{L})^{2}
 + (g_{V}^{L}- g_{A}^{L})^{2}(1-y)^{2}
- \frac{m_{e}y}{E_{\nu}}((g_{V}^{L})^{2} - (g_{A}^{L})^{2})]\}, \\
(\frac{d^{2} \sigma}{d y d \phi_{e'}})_{(S)}
&=& \mbox{}  B\{\frac{1}{8}y(y+2\frac{m_{e}}{E_{\nu}})
[ |g_{S}^{R}|^{2}(1+\mbox{\boldmath $\eta_{\nu}$}\cdot\hat{\bf q})]\}, \\
(\frac{d^{2} \sigma}{d y d \phi_{e'}})_{(V S)}
&=& \mbox{} B\{\sqrt{y(y+2\frac{m_{e}}{E_{\nu}})}[-\mbox{\boldmath
$\eta_{\nu}$}\cdot({\bf \hat{p}_{e'} \times \hat{q}})Im(g_{V}^{L}g_{S}^{R*})
+ (\mbox{\boldmath $\eta_{\nu}$}\cdot
{\bf \hat{p}_{e'}}) Re(g_{V}^{L}g_{S}^{R*})] \\
&& \mbox{} - y(1+\frac{m_{e}}{E_{\nu}}) (\mbox{\boldmath $\eta_{\nu}$}\cdot\hat{\bf q})
Re(g_{V}^{L}g_{S}^{R*})\},\nonumber
\eeq
\end{widetext}
\beq B & \equiv & \frac{E_{\nu}m_{e}}{(2\pi)^{2}}
\frac{G_{F}^{2}}{2}, \\
y & \equiv &
\frac{E_{e'}^{R}}{E_{\nu}}=\frac{m_{e}}{E_{\nu}}\frac{2cos^{2}\theta}
{(1+\frac{m_{e}}{E_{\nu}})^{2}-cos^{2}\theta}, \eeq where $y$-
the ratio of the kinetic energy of the recoil electron
$E^{R}_{e'}$  to the incoming neutrino energy $E_{\nu}$,
$\theta$- the angle between the direction of the outgoing
electron momentum  $ \hat{\bf p}_{e'}$  and the direction  of the
incoming neutrino momentum $\hat{\bf q}$ (recoil electron
scattering angle), $m_{e}$- the electron mass,$ \mbox{\boldmath
$\eta_{\nu}$}\cdot\hat{\bf q}$ - the longitudinal polarization of
the incoming neutrino,
$\phi_{e'}$ - the angle between the production plane and the
reaction plane.
\par
\begin{figure}
\includegraphics[width=11cm,angle=0]{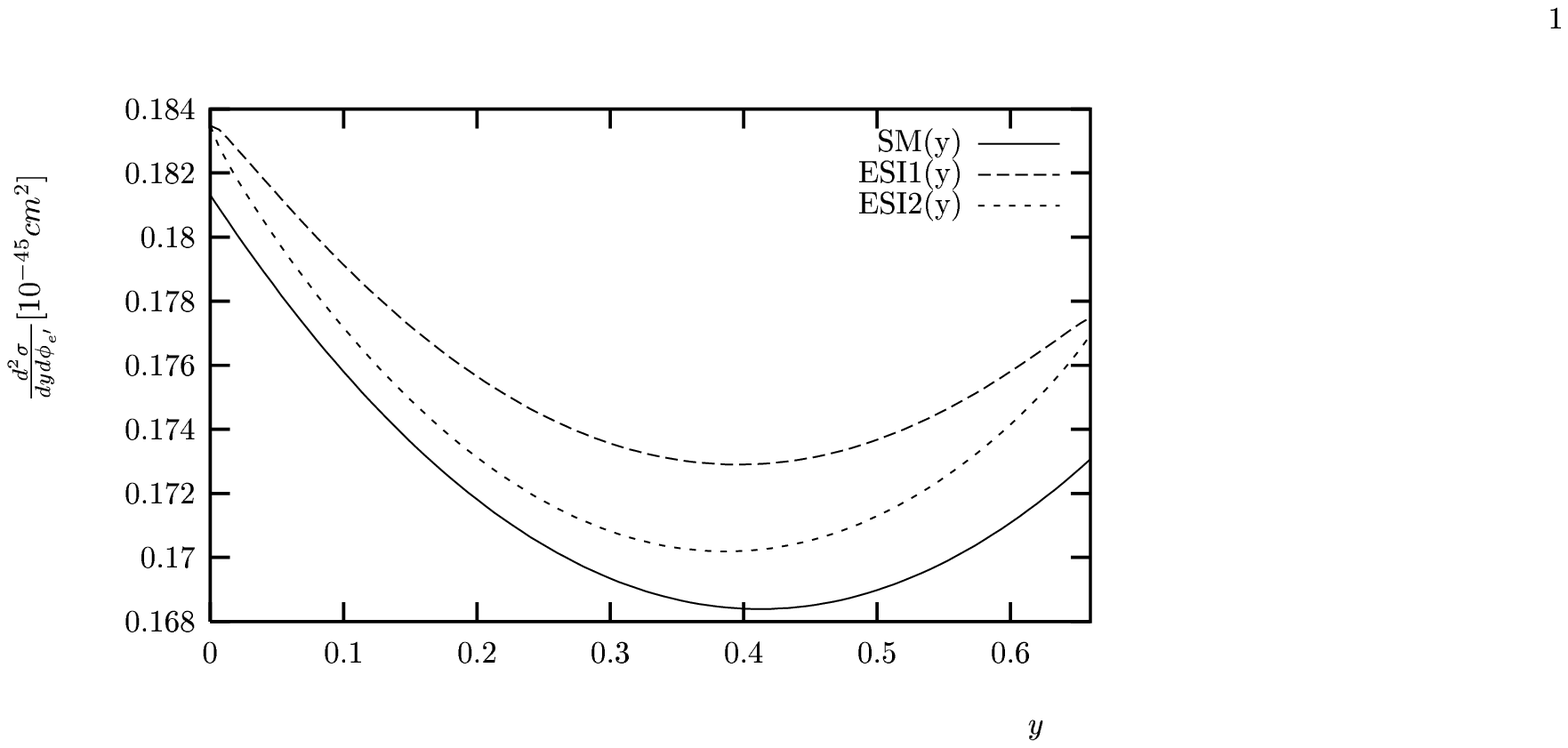}
\caption{Plot of the $\frac{d^{2} \sigma}{d y d \phi_{e'}}$ as a
function $y$ for; a) SM  with the left-handed neutrino (SM(y) -
solid line), b) the case of the exotic scalar interaction with the
right-handed neutrino for $\phi-\alpha=0$ (ESI1(y)- dashed line)
and  for $\phi-\alpha=\pi$ (ESI2(y) - dashed line), respectively.}
 \label{wykr}
\end{figure}

 It can be noticed that  the main
non-standard contributions to the laboratory differential cross
section come from the interference between the standard
left-handed vector $g_{V}^{L}$ coupling and exotic right-handed
scalar $g_{S}^{R}$ coupling, whose  occurrence  does not depend
explicitly on the neutrino mass. This interference may be
understood as the interference between the neutrino waves of the
same handedness for the the final neutrinos.
The similar regularity as for $\mu^{-}$-capture  appeared here \cite{Sobkow}.
The correlation $\mbox{\boldmath $\eta_{\nu}$}\cdot {\bf
\hat{p}_{e'}}$ proportional to $Re(g_{V}^{L}g_{S}^{R*})$ lies in
the  reaction plane spanned by the $ \hat{\bf p}_{e'}$ and $
\hat{\bf q}$, and it  includes both  $T$-even longitudinal and
transverse components of the initial neutrino polarization: \beq
(\mbox{\boldmath $\eta_{\nu}$}\cdot {\bf \hat{p}_{e'}})&=&
cos\theta(\mbox{\boldmath $\eta_{\nu}$}\cdot\hat{\bf q}) +
(\mbox{\boldmath $\eta_{\nu }^{' T}$}\cdot{\bf {p}_{e'}^{T}}),
 \eeq
where index ''T'' describes  transverse components
of the outgoing electron momentum  and of the incoming neutrino polarization,
respectively.
The other correlation  $\mbox{\boldmath $\eta_{\nu}$}\cdot({\bf
\hat{p}_{e'}\times \hat{q}})$ proportional to
$Im(g_{V}^{L}g_{S}^{R*})$ lies along the direction perpendicular
to the reaction plane and it includes only $T$-odd transverse
component of the initial neutrino polarization:
\beq
\mbox{\boldmath $\eta_{\nu}$}\cdot({\bf \hat{p}_{e'} \times
\hat{q}})&=& \mbox{\boldmath $\eta_{\nu}^{' T}$}\cdot({\bf
\hat{p}_{e'} \times \hat{q}}).
\eeq
It can be shown that in the
full interference term, Eq. (10), the contributions from the
longitudinal components of the neutrino polarization annihilate,
and in consequence one gives the interference  including only the
transverse components of the initial neutrino polarization, both
$T$-even and $T$-odd:
\beq \label{inter}
(\frac{d^{2} \sigma}{d y d \phi_{e'}})_{(VS)}  &=&
B\{\sqrt{\frac{m_{e}}{E_{\nu}}y[2-(2+\frac{m_{e}}{E_{\nu}})y]}\\
&& \times |g_{V}^{L}||g_{S}^{R}||\mbox{\boldmath $\eta_{\nu}^{'
T}$}|cos(\phi-\alpha)\}, \nonumber \eeq where $\alpha \equiv
\alpha_{V}^{L} - \alpha_{S}^{R} $ - the relative phase between
the  $g_{V}^{L}$ and  $g_{S}^{R}$ couplings, $\phi$ - the angle
between the reaction plane and the transverse neutrino
polarization vector and it is connected with the $\phi_{e'}$ in
the following way; $\phi=\phi_{0}-\phi_{e'}$, where
$\phi_{0}$ - the angle between the production plane and the
transverse neutrino polarization vector. It can be noticed that
the contribution from the interference between the $g_{V}^{L}$
and $g_{S}^{R}$ couplings, involving the transverse neutrino
polarization components, will be substantial at lower neutrino
energies $E_{\nu}\leq m_{e}$  but negligibly small at large
energies and vanishes for $\theta=0$ or $\theta=\pi/2$. The
occurrence of the interference term in the cross section depends
on the relative phase between the angle $\phi$ and phase $\alpha$
and does not vanish for $\phi - \alpha \not=\pi/2$. The situation
is illustrated in the Fig.\ref{wykr}, when $m_{e}/E_{\nu}=1$, $y
\in [0, 0.66]$, $\phi-\alpha=0$ (ESI1(y) - dashed line) and
$\phi-\alpha=\pi$ (ESI2(y) - dashed line), respectively.
We use the present model-independent upper limits on the
non-standard coupling constants \cite{Data} obtained from the
normal and inverse muon-decay, assuming the universality of weak
interaction. We take the upper bound on the $|\mbox{\boldmath
$\eta_{\nu}^{' T}$}|\leq 0.334$ for the neutrinos coming from the
muon-capture (however the phase $\alpha$ is still unknown). The
value $|\mbox{\boldmath $\eta_{\nu}^{' T}$}|=0.334$ is used to
get the upper limit on the expected effect from the right-handed
neutrinos in the cross section for the $(\nu_{\mu}e^{-})$
scattering. It means that the value of the longitudinal neutrino
polarization is equal to $\mbox{\boldmath $\eta_{\nu}^{l}$}\equiv
\mbox{\boldmath $\eta_{\nu}$}\cdot\hat{\bf q}=-0.943$.
The plot for the SM is made with the use of the present
experimental values for $g_{V}^{L}=-0.041\pm 0.015$,
$g_{A}^{L}=-0.507\pm 0.014$ \cite{Data}, $\mbox{\boldmath
$\eta_{\nu}$}\cdot\hat{\bf q}= -1$ and $ m_{e}/E_{\nu}=1$,
Fig.\ref{wykr} (SM(y) - solid line).
\par It is  known that in the SM the angular distribution of the
recoil electrons does not depend on the $\phi_{e'}$. It is
necessary to analyse  all the possible reaction planes
corresponding to  the given recoil electron scattering angle,
e.g. $\theta=\pi/3$,  $\theta=\pi/4$,  $\theta=2\pi/9$ (for
$E_{\nu}\leq m_{e})$ to verify if the azimuthal asymmetry in the
cross section appears. The regularity of this type  would
indicate the possible participation of the right-handed neutrinos
in the $(\nu_{\mu} e^{-})$ scattering.
\par The low-energy high-precision neutrino-electron scattering experiments
using the beta-radioactive intense and polarized neutrino source
could be used to search for  new effects  coming from the {\sl
right-handed neutrinos}, e. g. the Borexino neutrino experiment
with the chromium source  \cite{Miranda} (to be published).

\begin{acknowledgments}
I am greatly indebted to  Prof.\  S.\ Ciechanowicz for
many  useful and helpful discussions.
\end{acknowledgments}

\end{document}